\documentstyle[amssymb,12pt]{article}

\begin{document}

\def\baselinestretch{1.2}

\title{Integrable Equations on Time Scales}

\author{ Metin G{\" u}rses$^{(1)}$, Gusein Sh. Guseinov$^{(2)}$\\
 and Burcu Silindir$^{(1)}$  \\
{\small (1) Department of Mathematics, Faculty of Sciences}\\
{\small Bilkent University, 06800 Ankara, Turkey}\\
{\small (2) Department of Mathematics, Atilim University}\\
{\small 06836 Incek, Ankara, Turkey}\\}

\begin{titlepage}
\maketitle

\begin{abstract}
Integrable systems are usually given in terms of functions of
continuous variables (on ${\mathbb R}$), in terms of functions of
discrete variables (on ${\mathbb Z}$), and recently in terms of
functions of $q$-variables (on ${\mathbb K}_{q}$). We formulate
the Gel'fand-Dikii (GD) formalism on time scales by using the
delta differentiation operator and find more general integrable
nonlinear evolutionary equations. In particular they yield
integrable equations over integers (difference equations) and over
$q$-numbers ($q$-difference equations). We formulate the GD
formalism also in terms of  shift operators for all
regular-discrete time scales. We give a method allowing to
construct the recursion operators for integrable systems on time
scales. Finally, we give a trace formula on time scales and then
construct infinitely many conserved quantities (Casimirs) of the
integrable systems on time scales.

\end{abstract}

\end{titlepage}

\section*{1. Introduction}

Integrable systems are well studied and well understood in $1+1$
dimensions \cite{fok}-\cite{blaz1}. Here one of the dimensions
denotes the time (evolution) variable and the other one denotes
the space variable which is usually taken as  continuous. There
are also important examples where this variable takes values in
${\mathbb Z}$, i.e., integer values. In both cases the Gel'fand-
Dikii (GD) approach is quite effective. One can generate
hierarchies of integrable evolution equations, both on ${\mathbb
R}$ and on ${\mathbb Z}$ (see \cite{blaz1} for GD applications and
related references). In addition one can construct the conserved
quantities, Hamilton operators, and recursion operators.
Investigation of integrable systems on $q$-discrete intervals
started in \cite{fren}- \cite{ad}. They considered GD formalism on
${\mathbb K}_{q}$ and found $q$-integrable hierarchies including
the $q$-KdV  equation.

In this work we extend the Gel'fand- Dikii approach to time scales
where  ${\mathbb R}$, ${\mathbb Z}$, and ${\mathbb K}_{q}$ are
special cases. In the next section we give a brief review of time
scales calculus. See the references \cite{ah}-\cite{gus} for a
more detailed review of the subject . In GD formalism, in
obtaining integrable systems the essential tools are the
differential and shift operators and their inverses. For extending
the GD formulation to time scales we give necessary  means  to
construct in the sequel the algebra of pseudo $\Delta-$
differential operators and the algebra of shift operators. In
Section 3 we assume $\Delta$-differential Lax operators and derive
the $\Delta$-Burgers hierarchy with its recursion operator. We
present special cases of the Burgers equation for ${\mathbb
T}=h{\mathbb Z}$ and ${\mathbb T}={\mathbb K}_{q}$. In Section 4,
we consider the {\it regular} time scales where the inverse of
jump operators can be defined. Here we assume a pseudo
$\Delta$-differential algebra and give the corresponding GD
formulation. As an example we present a $\Delta$-KdV hierarchy. We
first find $n=1$ member of the hierarchy and write out it
explicitly for ${\mathbb T}={\mathbb R}, {\mathbb Z}, {\mathbb
K}_{q}$ and for ${\mathbb T}=(-\infty, 0) \cup {\mathbb K}_{q}$.
Then we give the $n=3$ member and call it as the $\Delta$-KdV
system. We call it $\Delta$-KdV equation, because the
corresponding Lax operator is a second order $\Delta$-differential
operator. It involves two fields $u$ and $v$, but the second field
$v$ can be expressed in terms of the first filed $u$. When
${\mathbb T}={\mathbb R}$, this system reduces to the standard KdV
equation. In Section 5, we consider the {\it regular-discrete}
time scales and introduce the algebra of shift operators on them
and give the corresponding GD formulation for all such time
scales.  Here several examples are presented. We first generalize
the examples of  discrete systems on ${\mathbb Z}$ given in
\cite{blaz1} (one field, two fields, and four fields examples in
\cite{blaz1}) to arbitrary discrete time scales. In all these
examples when ${\mathbb T}={\mathbb Z}$ we get the discrete
evolutions given in \cite{blaz1}. We construct the recursion
operators of these systems on time scales. We  generalize the
Frenkel's KdV system \cite{fren} introduced on ${\mathbb K}_{q}$
to arbitrary discrete time scales and we construct its recursion
operator. In this section, we finally give an example of the KP
hierarchy on discrete time scales. In Section 6, we extend the
standard way of constructing the conserved quantities of
integrable systems to time scales by introducing a {\it trace
form} on the algebra of $\Delta$-pseudo differential  operators.
The trace form introduced in this section reduces, in particular
cases, to the standard trace forms on ${\mathbb R}$ and  ${\mathbb
Z}$. In Appendix we give the recursion operators of two {\it four
fields systems} introduced in Section 5. We end up with a
conclusion.

\section*{2. Time Scale Calculus}

The time scale calculus is developed mainly to unify differential,
difference, and $q-$ calculus. A time scale $({\mathbb T})$ is an
arbitrary nonempty closed subset of the real numbers. The calculus
of time scales was initiated by Aulbach and Hilger
\cite{ah},\cite{hil} in order to create a theory that can unify
and extend discrete and continuous analysis. The real numbers
$({\mathbb R})$, the integers $({\mathbb Z})$, the natural numbers
$({\mathbb N})$, the non-negative integers $({\mathbb N}_{0})$,
the $h$-numbers $(h{\mathbb Z}=\{hk: k \in {\mathbb Z}\}$, where
$h>0$ is a fixed real number), and the $q-$ numbers $({\mathbb
K}_{q}=\,q^{\mathbb Z}\cup \{0\} \equiv \{ q^{k}: k \in {\mathbb
Z}\} \cup \{0\}$,\,\,where $q>1$ is a fixed real number) are
examples of time scales, as are $[0,1] \cup [2,3]$, $[0,1] \cup
{\mathbb N}$, and the Cantor set, where $[0,1]$ and $[2,3]$ are
real number intervals. In \cite{ah}, \cite{hil} Aulbach and Hilger
introduced also dynamic equations on time scales in order to unify
and extend the theory of ordinary differential equations,
difference equations, and quantum equations \cite{kac}
($h$-difference and $q$ -difference equations based on
$h$-calculus and $q$-calculus,respectively). For a general
introduction  to the calculus on time scales we refer the reader
to the textbooks by Bohner and Peterson \cite{boh1},\cite{boh2}.
Here we give only those notions and facts connected to time scales
which we need for our purpose in this paper.

Any time scale ${\mathbb T}$ is a complete metric space with the
metric (distance) $d(x,y)=|x-y|$ for $x,y \in {\mathbb T}$.
Consequently, according to the well-known theory of general metric
spaces, we have for ${\mathbb T}$ the fundamental concepts such as
open balls (intervals), neighborhood of points, open sets, closed
sets, compact sets, and so on. In particular, for a given number
$r> 0$, the $r$-neighborhood $U_{r}(x)$ of a given point $x \in
{\mathbb T}$ is the set of all points $y \in {\mathbb T}$ such
that $d(x,y) < r$. By a neighborhood of a point $x \in {\mathbb
T}$ is meant an arbitrary set in ${\mathbb T}$ containing an
$r$-neighborhood of the point $x$. Also we have for functions $f:
{\mathbb T} \rightarrow {\mathbb R}$ the concepts of the limit,
continuity, and properties of continuous functions on general
complete metric spaces (note that, in particular, any function $f:
{\mathbb Z} \rightarrow {\mathbb R}$ is continuous at each point
of ${\mathbb Z}$). The main task is to introduce and investigate
the concept of derivative for functions $f: {\mathbb T}
\rightarrow {\mathbb R}$. This proves to be possible due to the
special structure of the metric space ${\mathbb T}$. In the
definition of derivative, the so-called forward and backward jump
operators play special and important roles.

\vspace{0.3cm}

\noindent {\bf Definition 1}.\, {\it For $x \in {\mathbb T}$ we
define the forward jump operator $\sigma: {\mathbb T} \rightarrow
{\mathbb T}$ by
\begin{equation}
\sigma(x)=inf\, \{ y \in {\mathbb T}: y> x\},
\end{equation}
while the backward jump operator $\rho: {\mathbb T} \rightarrow
{\mathbb T}$ is defined by
\begin{equation}
\rho(x)=sup\, \{ y \in {\mathbb T}: y < x\}.
\end{equation}}

 \vspace{0.3cm}

\noindent In this definition we put in addition $\sigma(\max
{\mathbb T})=\max {\mathbb T}$ if there exists a finite
${\max{\mathbb T}}$, and $\rho(\min{\mathbb T})=\min{\mathbb T}$
if there exists a finite ${\min {\mathbb T}}$. Obviously both
$\sigma(x)$ and $\rho(x)$ are in ${\mathbb T}$ when $x \in
{\mathbb T}$. This is because of our assumption that ${\mathbb T}$
is a closed subset of ${\mathbb R}$.

Let $x \in {\mathbb T}$. If $\sigma(x)>x$, we say that $x$ is {\it
right-scattered}, while if $\rho(x)<x$ we say that $x$ is {\it
left-scattered}. Also, if $x < \max {\mathbb T}$ and
$\sigma(x)=x$, then $x$ is called {\it right-dense}, and if $x>
\min{\mathbb T}$ and $\rho(x)=x$, then $x$ is called {\it
left-dense}. Points that are right-scattered and left-scattered at
the same time are called {\it isolated}. Finally, the {\it
graininess functions} $\mu,\, \nu : {\mathbb T} \rightarrow
[0,\infty)$ are defined by
\begin{equation}
\mu(x)=\sigma (x)-x, ~~~\mbox{and}~~ \nu(x)=x-\rho(x)   ~~~~
\mbox{for all}~ x \in {\mathbb T}.
\end{equation}

\vspace{0.3cm}

\noindent {\bf Example 1}.\, If ${\mathbb T}={\mathbb R}$, then
$\sigma(x)=\rho(x)=x$ and $\mu(x)=\nu(x)=0$. If ${\mathbb
T}=h{\mathbb Z}$, then $\sigma(x)=x+h$, $\rho(x)=x-h$, and
$\mu(x)=\nu(x)=h$. On the other hand, if ${\mathbb T}={\mathbb
K}_{q}$ then we have
\begin{equation}
\sigma(x)=q\,x,~~~ \rho(x)=q^{-1}\,x , ~~~ \mu(x)=(q-1)x,~~
\mbox{and}~~~ \nu(x)= (1-q^{-1})\, x.
\end{equation}

Let ${\mathbb T}^{\kappa}$ denotes Hilger's above truncated set
consisting of ${\mathbb T}$ except for a possible left-scattered
maximal point. Similarly, ${\mathbb T}_{\kappa}$ denotes the below
truncated set obtained from ${\mathbb T}$ by deleting a possible
right-scattered minimal point.

\vspace{0.3cm}

\noindent {\bf Definition 2}.\, {\it Let $f: {\mathbb T}
\rightarrow {\mathbb R}$ be a function and $x \in {\mathbb
T}^{\kappa}$. Then the delta derivative of $f$ at the point $x$ is
defined to be the number $f^{\Delta}(x)$ (provided it exists) with
the property that for each $\varepsilon>0$ there exists a
neighborhood $U$ of $x$ in ${\mathbb T}$ such that
\begin{equation}\label{def2}
|f(\sigma(x))-f(y)-f^{\Delta}(x)[\sigma(x)-y]| \le \varepsilon
|\sigma(x)-y|,
\end{equation}
for all $ y \in U$.}

 \vspace{0.3cm}

\noindent {\bf Remark 1}.\, If $x \in {\mathbb T} \setminus
{\mathbb T}^{\kappa}$, then $f^{\Delta}(x)$ is not uniquely
defined, since for such a point $x$, small neighborhoods $U$ of
$x$ consist only of $x$ and besides we have $\sigma(x)=x$.
Therefore (\ref{def2}) holds for an arbitrary number
$f^{\Delta}(x)$. This is a reason why we omit a maximal
left-scattered point.

\vspace{0.3cm}

\noindent We have the following:\,\, (i)\, If $f$ is delta
differentiable at $x$, then $f$ is continuous at $x$. (ii)\, If
$f$ is continuous at $x$ and $x$ is right-scattered, then $f$ is
delta differentiable at $x$ with
\begin{equation}
f^{\Delta}(x)={f(\sigma(x))-f(x) \over \mu(x)}.
\end{equation}
(iii)\,If $x$ is  right-dense, then $f$ is delta differentiable at
$x$ iff the limit
\begin{equation}
\lim_{y \rightarrow x}\, {f(x)-f(y)) \over x-y}
\end{equation}
exists as a finite number. In this case $f^{\Delta}(x)$ is equal
to this limit. (iv)\, If $f$ is delta differentiable at $x$, then
\begin{equation}\label{for1}
f(\sigma(x))=f(x)+\mu(x)\, f^{\Delta}(x).
\end{equation}

\vspace{0.3cm}

\vspace{0.3cm}

\noindent {\bf Definition 3}.\,{\it If $x \in {\mathbb
T}_{\kappa}$, then we define the nabla derivative of $f: {\mathbb
T} \rightarrow {\mathbb R}$ at $x$ to be the number
$f^{\nabla}(x)$ (provided it exists) with the property that for
each $\varepsilon>0$ there is a neighborhood $U$ of $x$ in
${\mathbb T}$ such that
\begin{equation}
|f(\rho(x))-f(y)-f^{\nabla}(x)[\rho(x)-y]| \le \varepsilon
|\rho(x)-y|,
\end{equation}
for all $y \in U$.}

\vspace{0.3cm}

\noindent We have the following:~ (i) If $f$ is nabla
differentiable at $x$, then $f$ is continuous at $x$. (ii) If $f$
is continuous at $x$ and $x$ is left-scattered, then $f$ is nabla
differentiable at $x$ with

\begin{equation}
 f^{\nabla}(x)={f(x)-f(\rho(x)) \over \nu(x)}.
\end{equation}

\noindent (iii) If $x$ is left-dense, then $f$ is nabla
differentiable at $x$ if and only if the limit

\begin{equation}
\lim_{y \rightarrow x}\, {f(x)-f(y) \over x-y}
\end{equation}
exists as a finite number. In this case $f^{\nabla}(x)$ is equal
to this limit. (iv) If $f$ is nabla differentiable at $x$, then

\begin{equation}
f(\rho(x))=f(x)-\nu(x)\, f^{\nabla}(x).
\end{equation}

\noindent {\bf Example 2}.\,If ${\mathbb T}={\mathbb R}$, then
$f^{\Delta}(x)=f^{\nabla}(x)=f^{\prime}(x)$, the ordinary
derivative of $f$ at $x$. If ${\mathbb T}=h{\mathbb Z}$, then
\begin{equation}
f^{\Delta}(x)={f(x+h)-f(x) \over h}~~~~
\mbox{and}~~~~f^{\nabla}(x)={f(x)-f(x-h) \over h}.
\end{equation}
If ${\mathbb T}={\mathbb K}_{q}$, then
\begin{equation}
f^{\Delta}(x)={f(qx)-f(x) \over (q-1)x}
~~~\mbox{and}~~~f^{\nabla}(x)={f(x)-f(q^{-1}\,x) \over
(1-q^{-1})x},
\end{equation}
for all $x \ne 0$, and
\begin{equation}
f^{\Delta}(0)=f^{\nabla}(0)=\lim_{y \rightarrow 0}{f(y)-f(0) \over
y}
\end{equation}
provided that this limit exists.

\noindent Among the important properties of of the delta
differentiation on ${\mathbb T}$ we have the Leibnitz rule: If $f,
g: {\mathbb T} \rightarrow {\mathbb R}$ are delta differentiable
functions at $x \in {\mathbb T}^{\kappa}$, then so is their
product $f g$ and
\begin{eqnarray}
(f\,g)^{\Delta}(x)&=&f^{\Delta}(x)\, g(x)+f(\sigma(x)\,
g^{\Delta}(x),\label{for2}\\
&=&f(x)\, g^{\Delta}(x)+f^{\Delta}(x)\, g(\sigma(x)). \label{for3}
\end{eqnarray}
Also, if  $f, g: {\mathbb T} \rightarrow {\mathbb R}$ are nabla
differentiable functions at $x \in {\mathbb T}_{\kappa}$, then so
is their product $f g$ and
\begin{eqnarray}
(f\,g)^{\nabla}(x)&=&f^{\nabla}(x)\, g(x)+f(\rho(x)\,
g^{\nabla}(x),\label{for4}\\
&=&f(x)\, g^{\nabla}(x)+f^{\nabla}(x)\, g(\rho(x)). \label{for5}
\end{eqnarray}
In the next proposition we give a relationship between the delta
and nabla derivatives (see \cite{at}).

\vspace{0.3cm}

\noindent {\bf Proposition 4}.\, {\it (i) Assume that $f: {\mathbb
T} \rightarrow {\mathbb R}$ is delta differentiable on ${\mathbb
T}^{\kappa}$. Then $f$ is nabla differentiable at $x$ and
\begin{equation}\label{den1}
f^{\nabla}(x)=f^{\Delta}(\rho(x)),
\end{equation}
for $x \in {\mathbb T}_{\kappa}$ such that $\sigma(\rho(x))=x$.
If, in addition, $f^{\Delta}$ is continuous on ${\mathbb
T}^{\kappa}$, then $f$ is nabla differentiable at $x$ and
(\ref{den1}) holds for any $x \in {\mathbb T}_{\kappa}$.\\
(ii) Assume that  $f: {\mathbb T} \rightarrow {\mathbb R}$ is
nabla differentiable on ${\mathbb T}_{\kappa}$. Then $f$ is delta
differentiable at $x$ and
\begin{equation}\label{den2}
f^{\Delta}(x)=f^{\nabla}(\sigma(x)),
\end{equation}
for $x \in {\mathbb T}^{\kappa}$ such that $\rho(\sigma(x))=x$.
If, in addition, $f^{\nabla}$ is continuous on ${\mathbb
T}_{\kappa}$, then $f$ is delta differentiable at $x$ and
(\ref{den2}) holds for any $x \in {\mathbb T}^{\kappa}$.}

\vspace{0.3cm} Now we introduce the concept of integral for
functions $f: {\mathbb T} \rightarrow {\mathbb R}$.

\vspace{0.3cm}

\noindent {\bf Definition 5}.\,{\it A function $F: {\mathbb T}
\rightarrow {\mathbb R}$ is called a ${\Delta}$-antiderivative of
$f: {\mathbb T} \rightarrow {\mathbb R}$ provided
$F^{\Delta}(x)=f(x)$ holds for all $x$ in  ${\mathbb T}^{\kappa}$.
Then we define the $\Delta$-integral from $a$ to $b$ of $f$ by
\begin{equation}
\int_{a}^{b}\, f(x)\, \Delta\,x=F(b)-F(a)~~~\mbox{for all}~~ a,b
\in {\mathbb T}.
\end{equation}}

\vspace{0.3cm}

\noindent {\bf Definition 6}.\,{\it A function $\Phi: {\mathbb T}
\rightarrow {\mathbb R}$ is called a ${\nabla}$-antiderivative of
$f: {\mathbb T} \rightarrow {\mathbb R}$ provided
$\Phi^{\nabla}(x)=f(x)$ holds for all $x$ in  ${\mathbb
T}_{\kappa}$. Then we define the $\nabla$-integral from $a$ to $b$
of $f$ by
\begin{equation}
\int_{a}^{b}\, f(x)\, \nabla\,x=\Phi(b)-\Phi(a)~~~\mbox{for all}~~
a,b \in {\mathbb T}.
\end{equation}}

\noindent If $a,b \in {\mathbb T}$ with $a \le b$ we define the
closed interval $[a,b]$ in ${\mathbb T}$ by
\begin{equation}
[a,b]=\{x \in {\mathbb T}: a \le x \le b \}.
\end{equation}
Open and half-open intervals etc. are defined accordingly. Below
all our intervals will be time scale intervals

\vspace{0.3cm}

\noindent {\bf Example 3}.\, Let $a,b \in {\mathbb T}$ with $a<
b$. Then we have the following:\\
\noindent (i) If $f: {\mathbb T} \rightarrow {\mathbb R}$  then
\begin{equation}
\int_{a}^{b}\, f(x) \Delta\,x=\int_{a}^{b}\, f(x)\, \nabla
x=\int_{a}^{b}\, f(x) dx,
\end{equation}
where the integral on the right-hand side is the ordinary integral. \\
(ii) If $[a,b]$ consists of only isolated points, then
\begin{equation}
\int_{a}^{b}\, f(x) \Delta\,x=\sum_{x \in [a,b)}\, \mu(x)\,
f(x)~~~ \mbox{and}~~~\int_{a}^{b}\, f(x) \nabla\,x=\sum_{x \in
(a,b]}\, \nu(x)\,f(x).
\end{equation}
In particular, if ${\mathbb T}={\mathbb Z}$, then
\begin{equation}
\int_{a}^{b}\, f(x) \Delta\,x=\sum_{k=a}^{b-1} f(k)
~~~\mbox{and}\int_{a}^{b}\, f(x) \nabla\,x=\sum_{k=a+1}^{b} f(k).
\end{equation}
If ${\mathbb T}=h{\mathbb Z}$, then
\begin{equation}
\int_{a}^{b}\, f(x) \Delta\,x=h\sum_{x \in [a,b)} f(x)
~~~\mbox{and}\int_{a}^{b}\, f(x) \nabla\,x=h\, \sum_{x \in
(a,b]}^{b} f(x)
\end{equation}
and if ${\mathbb T}={\mathbb K}_{q}$, then
\begin{equation}
\int_{a}^{b}\, f(x) \Delta\,x=(1-q)\, \sum_{x \in [a,b)}\, xf(x)
~~~\mbox{and}~~~\int_{a}^{b}\, f(x) \nabla\,x=(1-q^{-1})\, \sum_{x
\in (a,b]}\,xf(x).
\end{equation}

The following relationship between the delta and nabla integrals
follows from Definitions 5 and 6  by using Proposition 4.

\vspace{0.3cm}

\noindent {\bf Proposition 7}. {\it If the function $f: {\mathbb
T} \rightarrow {\mathbb R}$ is continuous, then for all $a,b \in
{\mathbb T}$ with $ a<b$ we have
\begin{equation}\label{den21}
\int_{a}^{b} f(x) \Delta\,x=\int_{a}^{b} f(\rho(x)) \nabla\,x
~~~\mbox{and}~~~ \int_{a}^{b} f(x) \nabla\,x=\int_{a}^{b}
f(\sigma(x)) \Delta\,x.
\end{equation} }

\vspace{0.3cm}

\noindent Indeed, if  $F: {\mathbb T} \rightarrow {\mathbb R}$ is
a $\Delta$-antiderivative for $f$, then $F^{\Delta}(x)=f(x)$ for
all $x \in {\mathbb T}^{\kappa}$, and by Proposition 4 we have
$f(\rho(x))=F^{\Delta}(\rho(x))=F^{\nabla}(x)$ for all $x \in
{\mathbb T}_{\kappa}$, so that $F$ is  a $\nabla$-antiderivative
for $f(\rho(x))$. Therefore
\begin{equation}\label{den22}
\int_{a}^{b} f(\rho(x)) \nabla x=F(b)-F(a)=\int_{a}^{b} f(x)
\Delta x.
\end{equation}
From (\ref{for2})-(\ref{den2}) and (\ref{den21}) we have the
following integration by parts formulas: If the functions $f,g :
{\mathbb T} \rightarrow {\mathbb R}$ are delta and nabla
differentiable with continuous derivatives, then
\begin{eqnarray}
\int_{a}^{b} f^{\Delta}(x)\, g(x) \Delta x&=&f(x)\,
g(x)|_{a}^{b}-\int_{a}^{b} f(\sigma(x)) \, g^{\Delta}(x) \Delta x,
\\
\int_{a}^{b} f^{\nabla}(x)\, g(x) \nabla x&=&f(x)\,
g(x)|_{a}^{b}-\int_{a}^{b} f(\rho(x)) \, g^{\nabla}(x) \nabla x,
\\
\int_{a}^{b} f^{\Delta}(x)\, g(x) \Delta x&=&f(x)\,
g(x)|_{a}^{b}-\int_{a}^{b} f(x) \, g^{\nabla}(x) \nabla x,
\\
\int_{a}^{b} f^{\nabla}(x)\, g(x) \nabla x&=&f(x)\,
g(x)|_{a}^{b}-\int_{a}^{b} f(x) \, g^{\Delta}(x) \Delta x.
\end{eqnarray}

 For more general treatment of the delta integral on time
scales (Riemann and Lebesgue delta integrals on time scales) see
\cite{gus} and Chapter 5 of \cite{boh2}.

\section*{3. Burgers Equation on Time Scales}

\noindent The Gel'fand-Dikii approach is very effective in
studying the symmetries, bi-Hamiltonian formulation, and in
constructing the recursion operators of integrable nonlinear
partial differential equations. In this approach one takes the Lax
operator $L$ in an algebra like a differential or pseudo
differential algebra, a matrix algebra, a polynomial algebra, or
the Moyal algebra. In this section we take $L$ in the algebra of
delta differential operators.

Let ${\mathbb T}$ be a time scale. We say that a function $f:
{\mathbb T} \rightarrow {\mathbb R}$ is $\Delta$-smooth if it is
infinitely $\Delta$-differentiable (and hence infinitely
$\nabla$-differentiable).  By $\Delta$ we denote the {\it delta
differentiation operator} which assigns  to each
$\Delta$-differentiable function $f: {\mathbb T} \rightarrow
{\mathbb R}$ its delta derivative $\Delta(f)$ defined by
\begin{equation}
[\Delta(f)](x)=f^{\Delta}(x) ~~~~ \mbox{for}~~~ x \in {\mathbb
T}^{\kappa}.
\end{equation}
The {\it shift operator} $E$ is defined by the formula
\begin{equation}
(Ef)(x)=f(\sigma(x))
\end{equation}
for $x \in {\mathbb T}$, where $\sigma: {\mathbb T} \rightarrow
{\mathbb T}$ is the forward jump operator. It is convenient, in
the operator relations to denote  the delta differentiation
operator by $\delta$ rather than by $\Delta$. For example, $\delta
f$ will denote the composition (product) of the delta
differentiation operator $\delta$ and the operator of
multiplication by the function $f$. According to formula
(\ref{for2})) we have

\begin{equation}
\delta f=f^{\Delta}+E(f) \delta.
\end{equation}

\noindent Consider the $N$-th order $\delta$-differential operator
given by

\begin{equation}\label{lax1}
L=a_{N}\,{\delta}^{N}+a_{N-1}\,{\delta}^{N-1}+\cdots+a_{1}\,{\delta}+a_{0},
\end{equation}

\noindent where the coefficients $a_{i}$~ $(i=0,1,\cdots N)$ are
some $\Delta$-smooth functions of the variable $x \in {\mathbb
T}$. These functions are assumed to depend also on a continuous
variable $t \in {\mathbb R}$, however, we will not (for
simplicity) indicate explicitly the dependence on $t$.

\vspace{0.3cm}

\noindent {\bf Proposition 8}. {\it Let $L$ be given as in
(\ref{lax1}) and $A_{n}=(L^{n})_{>0}$ be the operator $L^{n}$
missing the $\delta^{0}$ term. Then the  Lax equation

\begin{equation}\label{lax2}
{dL \over dt_{n}}=[A_{n},L]=A_{n}\,L-L\, A_{n}
\end{equation}

\noindent for  $n=1,2, \cdots $ produces a consistent hierarchy of
coupled nonlinear evolutionary equations.}

\vspace{0.3cm}

\noindent {\bf Example 4. Burgers equation on time scale:}
 Let $L=v\,\delta+u$, where $u$ and $v$ are functions of $x \in {\mathbb T}$ and $t
 \in {\mathbb R}$. Then for an appropriate operator $A$ the Lax
 equation

\begin{equation}
{dL \over dt}=[A,L] \label{lax}
\end{equation}

\noindent defines a system  of two  differential equations for the
functions $u$ and $v$. We find the operator $A$ by using the
Gelfand-Dikii formalism. Let us start with the second power of $L$
and assume $A=(L^2)_{>0}$, where

\begin{equation}
L^2=v\, E(v)\, \delta^2+[v \, v^{\Delta}+v\, E(u)+uv]\, \delta+v\,
u^{\Delta}+u^2.
\end{equation}

\noindent We can assume $A=-(L^2)_{0}$ (the part of $-L^2$ without
the $\delta$ terms). With this choice, (\ref{lax}) gives

\begin{eqnarray}
{dv \over dt}&=&\mu\, v\, (v\, u^{\Delta}+u^2)^{\Delta}, \label{bur20}\\
{du \over dt}&=&v\,(v\,u^{\Delta}+ u^2)^{\Delta}, \, \label{bur21}
\end{eqnarray}
where $\mu(x)=\sigma(x)-x$ for $x \in {\mathbb T}$.

 \vspace{0.3cm}

\noindent Equations (\ref{bur20}) and (\ref{bur21}) given above
are not independent of each other. It is easy to see that $v=\mu\,
u+\lambda$, where $\lambda $ is an arbitrary real function
depending only on $x \in {\mathbb T}$. Then these two equations
reduce to a single equation, a Burgers equation on time scales,

\begin{equation}
{du \over dt}=(\mu\,u +\lambda)[u^2+(\mu u+\lambda)
u^{\Delta}]^{\Delta} .  \label{bur22}
\end{equation}
Let us present some special cases: {\bf (i)} When ${\mathbb
T}={\mathbb R}$ then $\mu=0$ and $\delta=D$, the usual
differentiation. Hence we can let $\lambda=1$ and (\ref{bur22})
reduces to the standard Burgers equation on ${\mathbb R}$.{\bf
(ii)} When ${\mathbb T}=h{\mathbb Z}$ then $\mu(m)=h$ and
$f^{\Delta}(m)= {1 \over h}[f(m+h)-f(m)]$ for any $f$. Then taking
$\lambda=0$ in (\ref{bur22})  we find

\begin{equation}
{du(m) \over dt}=u(m)\,u(m+h)\,[u(m+2h)-u(m)], \label{bur23}
\end{equation}
where $m \in h{\mathbb Z}$. The evolution equation given above in
(\ref{bur23}) represents a difference version of the Burgers
equation. {\bf (iii)} \,Let ${\mathbb T}=q^{\mathbb Z}$, where $q
\ne 1$ and $q>0$. Then we have $\mu(x)=(q-1)x$ and
$f^{\Delta}(x)={f(qx)-f(x) \over (q-1)x }$ and taking $\lambda=0$
we get from (\ref{bur22})

\begin{equation}
{du(x) \over dt}=u(x)u(qx)[u(q^2x)-u(x)].
\end{equation}
Taking $A_{n}=-(L^n)_{0}$ with $L$ given as in Example 4 we get a
hierarchy of evolution equations (Burgers hierarchy on time
scales) from

\begin{equation}
{dL \over dt_{n}}=-[(L^n)_{0}, v\, \delta+u]
\end{equation}

\noindent for all $n=1,2,3, \cdots$. Since $(L^n)_{0}$ is a scalar
function, letting $(L^n)_{0}=\rho_{n}$ we obtain

\begin{eqnarray}
{dv \over dt_{n}}&=&\mu\,v \, (\rho_{n})^{\Delta},\\
{du \over dt_{n}}&=&v\, (\rho_{n})^{\Delta},
\end{eqnarray}

\noindent where the first three $\rho_{n}$ are given by
\begin{eqnarray}
\rho_{1}&=&u,\\
\rho_{2}&=&v\,u^{\Delta}+u^2,\\
\rho_{3}&=&v E(v) u^{\Delta \Delta}+[vv^{\Delta}+v
E(u)+uv]u^{\Delta}+(vu^{\Delta}+u^2)u.
\end{eqnarray}
 The above hierarchy reduces to a single evolution equation with
$v=\mu\,u+\lambda$:

\begin{equation} \label{bur5}
{du \over dt_{n}}=(\mu\, u+\lambda)
(\tilde{\rho}_{n})^{\Delta},~~~n=1,2, \cdots
\end{equation}
where $\tilde{\rho}_{n}$ is equal to $\rho_{n}$ with $v=\mu\,
u+\lambda$. When ${\mathbb T}$ is a regular-discrete time scale,
the first three $\tilde{\rho}_{n}$ are given for $\lambda=0$ by

\begin{eqnarray}
\tilde{\rho}_{1}&=&u,\\
\tilde{\rho}_{2}&=&u\,E(u),\\
\tilde{\rho}_{3}&=&u\,E(u)\,E^2(u).
\end{eqnarray}

\vspace{0.3cm}

It is possible to construct the recursion operator ${\mathcal R}$
by using the Lax representation \cite{GKS}-\cite{GZh}. The
hierarchy satisfies a recursion relation like

\begin{equation}\label{rec}
{dL \over dt_{n+1}}=L\, {dL \over dt_{n}}+[R_{n},L], ~~~ n=1,2,
\cdots
\end{equation}

\noindent where $R_{n}$ is the remainder operator which has the
same degree as the Lax operator $L$. We shall construct this
operator for the Burgers equation with $\lambda=0$ on
regular-discrete time scales. Choosing $R_{n}=\alpha_{n}\, \delta$
we get (by choosing $v(x)=\mu(x)\, u(x)$)

\begin{equation}
{\mathcal R}=u\, E+[u\,(E(u)-u)]\, (E-1)^{-1}\, {E \over E(u)}.
\end{equation}

\noindent One can generate the hierarchy (\ref{bur5}) by
application of the recursion operator ${\mathcal R}$ to the lowest
order symmetry $u_{1}=u\, (E(u)-u)$:

\begin{equation}
{du \over dt_{n}}={\mathcal R}^{n-1}\, u_{1}, ~~~ n=1,2, \cdots .
\end{equation}

\vspace{0.3cm}

\section*{4. Algebra of Pseudo Delta Differential Operators on Regular
Time Scales}

Let us define the notion of regular time scales.

\vspace{0.3cm}

\noindent {\bf Definition 9}.\, {\it We say that a time scale
${\mathbb T}$ is regular if the following two conditions are
satisfied simultaneously :
\begin{eqnarray}
&&(i)~~ \sigma(\rho(x))=x~~ \mbox{for all}~~ x \in {\mathbb T}~~\mbox{and} \label{den3}\\
&&(ii)~~ \rho(\sigma(x))=x~~ \mbox{for all} ~~x \in {\mathbb T},
\label{den4}
\end{eqnarray}
where $\sigma$ and $\rho$ denote the forward and backward jump
operators, respectively.}

\vspace{0.3cm}

From (\ref{den3}) it follows that the operator $\sigma: {\mathbb
T} \rightarrow {\mathbb T}$ is "onto" while (\ref{den4}) implies
that $\sigma$ is "one-to-one". Therefore $\sigma$ is invertible
and $\sigma^{-1}=\rho$. Similarly, the operator $\rho: {\mathbb T}
\rightarrow {\mathbb T}$ is invertible and $\rho^{-1}=\sigma$ if
${\mathbb T}$ is regular.

Let us set $x_{*}=\min {\mathbb T}$ if there exists a finite $\min
{\mathbb T}$, and set $x_{*}=-\infty$ otherwise. Also set
$x^{*}=\max {\mathbb T}$ if there exists a finite $\max {\mathbb
T}$, and $x^{*}=\infty$ otherwise. It is not difficult to see that
the following statement holds.

\vspace{0.3cm}

\noindent {\bf Proposition 10}. {\it A time scale ${\mathbb T}$ is
regular
if and  only if the following two conditions hold: \\
(i) \, The point $x_{*}={\mathbb T}$ is right-dense and the point
$x^{*}=\max {\mathbb T}$ is left-dense.\\
(ii) \, Each point of ${\mathbb T} \setminus \{x_{*},x^{*}\}$ is
either two-sided dense or two-sided scattered.}

\vspace{0.3cm}

 \noindent
 In particular, ${\mathbb R}$, $h\,{\mathbb Z}$, and ${\mathbb
 K}_{q}$ are regular time scales, as are $[0,1]$, $
 [-1,0] \cup \{{1 \over k}: k \in {\mathbb N}\} \cup \{{k \over
 k+1}: k \in {\mathbb N} \} \cup [1,2]$, and $(-\infty , 0] \cup
 \{{1 \over k}: k \in {\mathbb N}\} \cup \{2k: k \in {\mathbb
 N}\}$, where $[-1,0], [0,1], [1,2], (-\infty, 0]$ are real line
 intervals.

If $f: {\mathbb T} \rightarrow {\mathbb R}$ is a function we
define the functions $f^{\sigma}: {\mathbb T} \rightarrow {\mathbb
R}$ and $f^{\rho}: {\mathbb T} \rightarrow {\mathbb R}$ by
\begin{equation}
f^{\sigma}(x)=f(\sigma(x))~~ \mbox{and}~~ f^{\rho}(x)=f(\rho(x))~~
\mbox{for all}~~ x \in {\mathbb T}.
\end{equation}
Defining the shift operator $E$ by the formula $E\,f=f^{\sigma}$
we have
\begin{equation}
(E\, f)(x)=f^{\sigma}(x)=f(\sigma(x))~~ \mbox{for all} ~~ x \in
{\mathbb T}.
\end{equation}
The inverse $E^{-1}$ exists only in case of regular time scales
and is defined by
\begin{equation}
(E^{-1}\,f)(x)=f(\sigma^{-1}(x))=f(\rho(x))~~\mbox{for all} ~~ x
\in {\mathbb T}.
\end{equation}
In the operator relations, for convenience, we will  denote the
shift operator by ${\cal E}$ rather than by $E$. For example,
${\cal E}\,f$ will denote the composition (product) of the shift
operator ${\cal E}$ and the operator of multiplication by the
function $f$. Obviously, for any integer $m \in {\mathbb Z}$, we
have
\begin{equation}
{\cal E}^{m}\, f=(E^{m}\, f)\, {\cal E}^{m}.
\end{equation}

Remember that  $\delta$ denotes  the delta differentiation
operator acting in the operator relations by $\delta
f=f^{\Delta}+E(f) \delta$. The following proposition is an
immediate consequence of the the formulas (\ref{for1}),
(\ref{for2}).

\vspace{0.3cm}

\noindent {\bf Proposition 11}.\, {\it The operator formulas
\begin{eqnarray}
{\cal E}=I+\mu\, \delta,~~ \mbox{and}\\
\delta\, f=f^{\Delta}+E(f)\, \delta \label{del68}
\end{eqnarray}
hold, where the function $\mu: {\mathbb T} \rightarrow {\mathbb
R}$ is defined by $\mu(x)=\sigma(x)-x$ for all $x \in {\mathbb
T}$, and $I$ denotes the identity operator.}

\noindent In this section we will assume that all our considered
functions from ${\mathbb T}$ to ${\mathbb R}$ are $\Delta$-smooth
and tend to zero sufficiently rapidly together with their
$\Delta$-derivatives as $x$ goes to $x_{*}$ or $x^{*}$, where
$x_{*}=\min {\mathbb T}$ if there exists a finite $\min {\mathbb
T}$ and  $x_{*}=-\infty$ otherwise, $x^{*}=\max {\mathbb T}$ if
there exists a finite $\max {\mathbb T}$ and $x^{*}=\infty$
otherwise. The inverse operator $\delta^{-1}$ exists on such
functions. If $g: {\mathbb T} \rightarrow {\mathbb T}$ is such a
function, then

\begin{equation}
[\Delta^{-1}(g)](x)=\int_{x_{*}}^{x}\, g(y) \, \Delta y .
\end{equation}

\vspace{0.3cm}

\noindent {\bf Proposition 12}.\, {\it Let $f: {\mathbb T}
\rightarrow {\mathbb R}$ be a $\Delta$-smooth function such that
$f$ and all its $\Delta$-derivatives vanish rapidly at $x_{*}$ and
$x^{*}$. Then the operator $\delta^{-1}\, f$ being the composition
(product) of $\delta^{-1}$ and $f$ has the form of the formal
series in powers of $\delta^{-1}$
\begin{equation}
\delta^{-1}\, f=\alpha_{0}\, \delta^{-1}+\alpha_{1}\,
\delta^{-2}+\cdots
\end{equation}
where $\alpha_{0}=E^{-1}\,f$, and $\alpha_{k}=(-1)^k\,
(E^{-1}\,f)^{\nabla^{k}}$ for $k=1,2,\cdots$}.

\vspace{0.3cm}

\noindent {\bf Proof}.\, Multiplying (\ref{del68}) on the left and
right by $\delta^{-1}$ we obtain

\begin{equation}
\delta^{-1}\, E(f)=f \delta^{-1}-\delta^{-1} \, f^{\Delta}\,
\delta^{-1}.
\end{equation}
Replacing here $f$ by $E^{-1} f$ we get
\begin{eqnarray}\label{eq00}
\delta^{-1}\, f=(E^{-1}\, f)\, \delta^{-1}-\delta^{-1}
(E^{-1}\,f)^{\Delta}\, \delta^{-1}.
\end{eqnarray}
Further, applying this rule to the function $(E^{-1} f)^{\Delta}$
and taking into account that by Proposition 4(i)

\begin{equation}
E^{-1}\,(E^{-1}\, f)^{\Delta}=(E^{-1})^{\nabla},
\end{equation}
we find

\begin{equation}\label{eq01}
\delta^{-1}\,(E^{-1} f)^{\Delta}=(E^{-1}\,f)^{\nabla}
\delta^{-1}-\delta^{-1}\, ((E^{-1} f)^{\nabla})^{\Delta}\,
\delta^{-2}.
\end{equation}
Substituting this into the second term on the right-hand side of
(\ref{eq00}) we obtain
\begin{equation}
\delta^{-1}\, f=(E^{-1}\, f) \delta^{-1}-(E^{-1} f)^{\nabla}\,
\delta^{-2}+\delta^{-1} \, ((E^{-1} f)^{\Delta})^{\nabla}\,
\delta^{-2}.
\end{equation}
Continuing this procedure repeatedly we arrive at the statement of
the proposition.

\vspace{0.3cm}

\noindent {\bf Definition 13}.\, {\it By $\Lambda$ we denote the
algebra of pseudo delta differential operators. Any operator $K
\in \Lambda$ of order $k$ has the form

\begin{equation}\label{ell}
K=\sum_{\ell=-\infty}^{k} a_{\ell}\, \delta^{\ell}
\end{equation}

\vspace{0.3cm}

\noindent where $a_{\ell}$'s are $\Delta$-smooth functions of $x
\in {\mathbb T}$. For $K$ given by (\ref{ell}) we will use the
following notations:

\begin{equation}
K_{\ge 0}=\sum_{\ell=0}^{k}\, a_{\ell}\, \delta^{\ell},~~
\mbox{and} ~~ K_{< 0}= \sum_{-\infty}^{-1}\, a_{\ell}\,
\delta^{\ell}.
\end{equation}
}

\vspace{0.3cm}

\noindent As an example we let

\begin{equation} \label{lax4}
L=a_{N}\, \delta^{N}+a_{N-1}\, \delta^{N-1}+ \cdots +a_{1}\,
\delta+a_{0}
\end{equation}
where $a_{i} (i=0,1, \cdots,N)$ are some $\Delta$-smooth functions
on ${\mathbb T}$. Then we have

\vspace{0.3cm}

\noindent {\bf Proposition 16}.\, {\it Let $L$ be given in
(\ref{lax4}). For each fixed $N$ the Lax equation

\begin{equation}\label{lax5}
{dL \over dt_{n}}=[A_{n},L], ~~~~ A_{n}=(L^{n \over N})_{ \ge 0},
\end{equation}
for $n=1,2, \cdots $ not divisible by $N$, produces a (consistent)
hierarchy of evolution equations (a KdV hierarchy on time scales)
.}

\vspace{0.3cm}

\noindent {\bf Proof}.\, Since $(L^{n \over N})_{\ge 0}= L^{n
\over N}-(L^{n \over N})_{<0}$, we get

\begin{equation}\label{gd1}
{dL \over dt_{n}}=[(L^{n \over N})_{ \ge 0},L]=-[L^{n \over N})_{
<0},L].
\end{equation}
Evidently the commutator $[(L^{n/N})_{\ge 0},L]$ involves only
nonnegative powers of $\delta$, while the commutator
$[(L^{n/N})_{<0},L]$ has the form $\sum_{j=-\infty}^{N-1}\,
b_{j}\, \delta^{j}$. Therefore, we get by (\ref{gd1}) that, for
all $n$ not divisible by $N$, (\ref{lax5}) produces nontrivial
consistent $N+1$-number of evolutionary coupled
$\Delta$-differential equations for $a_{i},~ i=0,1, \cdots , N$.
Note that $a_{N}$ turns out to be a fixed (i.e., time independent)
function of $x$.

\vspace{0.3cm}

\noindent {\bf Example 5}.\, A {\bf KdV hierarchy on time scales}.
Let
\begin{equation}\label{lax55}
L=\delta^2+v \delta+u,
\end{equation}
where $u$ and $v$ are $\Delta$-smooth functions. It is
straightforward to find that
\begin{equation}
L^{1/2}=\delta+\alpha_{0}+\alpha_{1}\, \delta^{-1}+\alpha_{2}\,
\delta^{-2}+\cdots
\end{equation}
where
\begin{eqnarray}
E(\alpha_{0})+\alpha_{0}=v, \label{alfa1}\\
E(\alpha_{1})+\alpha_{1}+(\alpha_{0})^{\Delta}+(\alpha_{0})^{2}=u, \label{alfa2}\\
E(\alpha_{2})+\alpha_{2}+\alpha_{1}\,
E^{-1}(\alpha_{0})+(\alpha_{1})^{\Delta}=0. \label{alfa3}
\end{eqnarray}
Choosing $n=1,3, \cdots$ we get the members of the KdV hierarchy.

\vspace{0.3cm} \noindent {\bf (1)}. Let $n=1$.  Then Lax equation
(\ref{lax5}) becomes
\begin{equation}
{dv \over dt}\, \delta+{du \over dt}=[(L^{1/2})_{ \ge 0}, L]
\end{equation}
and gives  coupled equations for $u$ and $v$
\begin{eqnarray}
{du \over dt}&=&u^{\Delta}-v\,(\alpha_{0})^{\Delta}-(\alpha_{0})^{\Delta \Delta},
\label{kdv1}\\
{dv \over dt}&=&v^{\Delta}+E(u)-u-v\,[E(\alpha_{0})-\alpha_{0}] \nonumber\\
&&-E(\alpha_{0}^{\Delta})-E(\alpha_{0})^{\Delta}, \nonumber\\
&=& \mu
(u^{\Delta}-v\,(\alpha_{0})^{\Delta}-(\alpha_{0})^{\Delta\Delta}).
\label{kdv2}
\end{eqnarray}
Comparing the above equations we get
\begin{equation}
{dv \over dt}-\mu {du \over dt}=0,
\end{equation}
and therefore
\begin{equation}
v=\mu\, u+\lambda,
\end{equation}
where $\lambda$ is an arbitrary real function depending only on $x
\in {\mathbb T}$. Thus, two equations (\ref{kdv1}) and
(\ref{kdv2}) reduce to the following  single equation
\begin{equation}
{du \over dt}=u^{\Delta}-(\mu \, u+\lambda)\,
(\alpha_{0})^{\Delta}-(\alpha_{0})^{\Delta \Delta}, \label{kdv3}
\end{equation}
where $\alpha_{0}$ is expressed, according to (\ref{alfa1}), from
\begin{equation}
E(\alpha_{0})+\alpha_{0}=\mu \, u+\lambda. \label{kdv4}
\end{equation}
If we take $\lambda=0$, then (\ref{kdv3}) and (\ref{kdv4})
become
\begin{eqnarray}
&&{du \over dt}=u^{\Delta}-\mu \, u\,
(\alpha_{0})^{\Delta}-(\alpha_{0})^{\Delta \Delta}, \label{kdv5}\\
&&E(\alpha_{0})+\alpha_{0}=\mu \, u. \label{kdv6}
\end{eqnarray}
We shall now give $\alpha_{0}$, for illustration,  for particular
cases of ${\mathbb T}$:

\vspace{0.3cm} \noindent
 {\bf (i)} In the case ${\mathbb
T}={\mathbb R}$ we have $\mu=0$ and (\ref{kdv6}) gives
$\alpha_{0}=0$ and (\ref{kdv5}) becomes
\begin{equation}
{du \over dt}={du \over dx},
\end{equation}
which is a linear equation explicitly solvable:
\begin{equation}
u(x,t)=\varphi(x+t),
\end{equation}
where $\varphi$ is an arbitrary differentiable function.

\vspace{0.3cm} \noindent {\bf (ii)} In the case ${\mathbb
T}={\mathbb Z}$ we have $\mu=1$ and (\ref{kdv6}) is satisfied by
\begin{equation}
\alpha_{0}(n)=-\sum_{k=-\infty}^{n-1}\, (-1)^{n+k}\, u(k), ~~ n
\in {\mathbb Z}
\end{equation}
and therefore the equation (\ref{kdv5}) becomes
\begin{equation}
{du(n) \over dt}=-u^2(n)+2u(n)+2(-1)^n\, [2+u(n)]\,
\sum_{k=-\infty}^{n-1}\, (-1)^{k}\, u(k),
\end{equation}
for $n \in {\mathbb Z}$.

\vspace{0.3cm}

\noindent {\bf (iii)} In the case ${\mathbb T}={\mathbb K}_{q}$ we
have $\mu(x)=(q-1) x$ and (\ref{kdv6}) is satisfied by
$\alpha_{0}(0)=0$ and
\begin{equation}\label{alfa51}
\alpha_{0}(x)=-(q-1)\sum_{y \in (0, q^{-1}x]}\,
(-1)^{log_{q}(xy)}\, y u(y)
\end{equation}
for $x \in {\mathbb K}_{q}$ and $x \ne 0$. Substituting
(\ref{alfa51}) into (\ref{kdv5}) we can get an evolution equation
for $u$.

\vspace{0.3cm} \noindent {\bf (iv)}\, Let ${\mathbb T}=(-\infty,
0) \cup {\mathbb K}_{q}=(-\infty,0] \cup q^{\mathbb Z}$. In this
case $\mu(x)=0$ if $x \in (-\infty, 0]$ and $\mu(x)=(q-1) x$ if $x
\in q^{\mathbb Z}$. The equation (\ref{kdv6}) is satisfied by the
function  $\alpha_{0}$ given by

\begin{equation}
\alpha_{0}(x)=\left \{ \begin{array}{ll} 0 & x \in (-\infty, 0]\\
-(q-1)\Sigma_{y \in (0, q^{-1}x]}\, (-1)^{log_{q}(xy)}\, y u(y)&
x\in q^{\mathbb Z}
\end{array} \right.
\end{equation}
Therefore (\ref{kdv5}) will yield an evolution equation coinciding
on $(-\infty, 0]$ and $q^{\mathbb Z}$ with the evolution equations
described in the examples (i) and (iii), respectively. Now an
essential complementary point is that the solution {\it u} must
satisfy at $x=0$ the smoothness conditions
\begin{equation}
u(0^{-})=u(0^{+}),~~~ u^{\prime}(0^{-})=u^{\Delta}(0^{+}).
\end{equation}

\vspace{0.3cm} \noindent {\bf (2)}. Letting $n=3$, first we get

\begin{equation}
L^{3/2}=\delta^{3}+p\, \delta^{2}+q\, \delta+r+ (\mbox{terms with
negative powers of } \delta)
\end{equation}
where
\begin{eqnarray}
p&=&\alpha_{0}+E(v), \\
q&=&v^{\Delta}+E(u)+\alpha_{0}\,v+\alpha_{1}, \\
r&=&u^{\Delta}+\alpha_{0}\, u+\alpha_{1}\, E^{-1}(v)+\alpha_{2},
\end{eqnarray}
and the Lax equation
\begin{equation}
{dv \over dt}\, \delta+{du \over dt}=[(L^{3/2})_{ \ge 0}, L],
\end{equation}
gives the coupled equations for $u$ and $v$
\begin{eqnarray}
{du \over dt}&=&u^{\Delta \Delta \Delta }+p\,u^{\Delta \Delta}+q\,
u^{\Delta}-r^{\Delta \Delta}-v\,r^{\Delta}, \\
{dv \over dt}&=&v^{\Delta \Delta \Delta}+E(u^{\Delta
\Delta})+(E(u^{\Delta}))^{\Delta}+E(u)^{\Delta \Delta}+p\,
[v^{\Delta
\Delta}+E(u^{\Delta}) \nonumber\\
&&+E(u)^{\Delta}] +q\,(v^{\Delta}+E(u)-u)+rv-q^{\Delta
\Delta}-E(r^{\Delta})\nonumber\\
&&-E(r)^{\Delta}-v\,q^{\Delta}-v E(r).
\end{eqnarray}

 \vspace{0.3cm}

 \noindent
 As in the first member of the hierarchy ($n=1$ case), the above
 $\Delta$-KdV equations reduce to
 a single equation for the function $u$. Below in Corollary 23 we found
 that $v=\mu(x) u+\lambda(x)$. Letting $\lambda=$ constant we get

 \begin{equation}
{du \over dt}=u^{\Delta \Delta \Delta }+p\,u^{\Delta \Delta}+q\,
u^{\Delta}-r^{\Delta \Delta}-v\,r^{\Delta}.
\end{equation}
It is possible to write the above equation more explicitly in
terms of $u$ for ${\mathbb T}={\mathbb R}$, ${\mathbb T}={\mathbb
Z}$, and for ${\mathbb T}={\mathbb K}_{q}$ but they are quite
lengthy. For the discrete case we give a KdV hierarchy in Example
8, next section.

\section*{5. Shift Lax Operators on Regular-Discrete Time Scales}

Let ${\mathbb T}$ be a time scale. Let us set $x_{*}=\min {\mathbb
T}$ if there exists a finite $\min {\mathbb T}$ and
$x_{*}=-\infty$ otherwise. Also set $x^{*}=\max {\mathbb T}$ if
there exists a finite $\max {\mathbb T}$ and $x^{*}=\infty$
otherwise. We will briefly write $x_{*}=\min {\mathbb T}$ and
$x^{*}=\max {\mathbb T}$.

\vspace{0.3cm}

\noindent { \bf Definition 17}.\,{\it We say that a time scale
${\mathbb T}$ is regular-discrete if the following two conditions
are satisfied:\\
(i) The point $x_{*}$ is right-dense and the point $x^{*}$ is
left-dense.\\
(ii) Each point of ${\mathbb T}\backslash \{x_{*},x^{*}\}$ is
two-sided scattered (isolated)}.

\vspace{0.3cm}

The shift operator $E$ is defined on functions $f: {\mathbb T}
\rightarrow {\mathbb R}$ by the formula

\begin{equation}
(Ef)(x)=f(\sigma(x)), ~~~ \mbox{for} ~~ x \in {\mathbb T}
\end{equation}

\noindent where $\sigma: {\mathbb T} \rightarrow {\mathbb T}$ is
the forward jump operator.

 In this section we deal
only with regular-discrete time scales ${\mathbb T}$. For such
time scales ${\mathbb T}$ we have
\begin{equation}
\mu(x)=\sigma(x)-x \ne 0,~~ \mbox{for all}~ x \in {\mathbb
T}\backslash \{x_{*},x^{*}\}
\end{equation}
and, therefore, on functions given on ${\mathbb T} \backslash
\{x_{*},x^{*}\}$ we have the operator relationship
\begin{equation}
\delta={1 \over \mu}\,({\mathcal E}-1).
\end{equation}
All our functions will be assumed to be defined on  ${\mathbb T}
\backslash \{x_{*},x^{*}\}$ and tends to zero sufficiently rapidly
as $x$ goes  to $x_{*}$ or $x^{*}$.

This shift operator ${\mathcal E}$, should be quite useful in the
application of the Gel'fand-Dikii formalism. The reason is that
for any integer $m$ we have the simple product rule
\begin{equation}\label{epsil}
{\mathcal E}^{m} u=(E^{m} u) \,{\mathcal E}^{m}.
\end{equation}
Hence, for regular-discrete time scales, we can define an algebra
of ${\mathcal E}$ operators.

\vspace{0.3cm}

\noindent {\bf Definition 18}. {\it An algebra ,
$\Lambda_{\epsilon}$, of  ${\mathcal E}$ operators satisfying the
operator equation (\ref{epsil}) is defined as follows: Any
operator $K$  in $\Lambda_{\epsilon}$ with degree $k$ is of the
form

\begin{equation}
K=\sum_{-\infty}^{k}\, a_{\ell}\, {\mathcal E}^{\ell}
\end{equation}
where $a_{\ell}$ are functions  of $x \in {\mathbb T}$ that depend
also on $t \in {\mathbb R}$.}

\vspace{0.3cm}

\noindent
 Hence we can form Lax operators in $\Lambda_{\epsilon}$,
 and produce integrable equations on regular-discrete time
scales. Following \cite{blaz1} we obtain two classes of Lax
representations.

\vspace{0.3cm}

\noindent {\bf Proposition 19}.\, {\it The Lax equation

\begin{equation}\label{lax19}
{dL \over dt_{\ell}}=[(L^{\ell})_{ \ge k},L],~~~~k=0,1
\end{equation}
produces consistent hierarchy of equations for $\ell=1,2,\cdots$
with the following suitable Lax operators

\begin{eqnarray}
L&=&{\mathcal E}^{\alpha+n}+u_{\alpha+n-1}\,{\mathcal
E}^{\alpha+n-1}+\cdots+u_{\alpha} {\mathcal E}^{\alpha}, \\
L&=&v_{\alpha+n}\,{\mathcal
E}^{\alpha+n}+v_{\alpha+n-1}\,{\mathcal
E}^{\alpha+n-1}+\cdots+v_{\alpha+1}\, {\mathcal
E}^{\alpha+1}+{\mathcal E}^{\alpha},
\end{eqnarray}
for $k=0$ and $k=1$, respectively. Here $u_{i}$ and $v_{i}$ are
functions  defined on ${\mathbb T}$ and the integer $\alpha$ is
restricted to satisfy the inequality $-n < \alpha \le -1$}.

\vspace{0.3cm}

\noindent {\bf Remark.} Lax operators above and the following
examples are given on any regular-discrete time scale ${\mathbb
T}$ (we can take in particular ${\mathbb T}={\mathbb Z}$ or
${\mathbb K}_{q}$). This means that for any function $u$ on such a
time scale $E(u)=u(\sigma(x))$ where $\sigma$ is the jump operator
defined in the second section. Hence our examples and results
should be considered as more general than those considered in
\cite{blaz1}. In the case of Ref.\cite{blaz1} time scale is just
the integers (${\mathbb T}={\mathbb Z}$) where $E(u(n))=u(n+1)$

\vspace{0.3cm}

 \vspace{0.3cm}

\noindent {\bf Example 6}.\, {\bf Two field equations}. Let $k=0$,
$\alpha=-1$ and
\begin{equation}
L=u_{-1}\, {\mathcal E}^{-1}+u_{0}+{\mathcal E} \equiv v\,
{\mathcal E}^{-1}+u+ {\mathcal E}.
\end{equation}
Then we find
\begin{eqnarray}
\ell=1~~~~{dv \over dt_{1}}&=&v(u-E^{-1}(u)), \\
{du \over dt_{1}}&=&E(v)-v,
\end{eqnarray}
\begin{eqnarray}
\ell=2~~~~{dv \over dt_{2}}&=&u^2\,v+E(v)\,v-vE^{-1}(v)-v\,E^{-1}(u^2), \\
{du \over dt_{2}}&=&u\,E(v)+E(u)\,E(v)-v\,E^{-1}(u)-u\,v,
\end{eqnarray}
\begin{eqnarray}
\ell=3~~~~{dv \over dt_{3}}&=&uv^2+u^3 v-vE^{-1}(v)E^{-2}(u)-2vE^{-1}(u)E^{-1}(v)- \nonumber\\
&&v E^{-1}(u^3)+2uvE(v)-v^2E^{-1}(u)+v E(u)E(v),\\
{du \over
dt_{3}}&= &E(v)[u^2+u E(u)+E(u^2)+E(v)+(E^2(v))]- \nonumber\\
&&v[E^{-1}(v)+E^{-1}(u^2)+uE^{-1}(u)+u^2+v].
\end{eqnarray}
This is a Toda hierarchy on discrete time scales. The recursion
relation between the $n+1$-th and  $n$-th elements of the
hierarchy is given by

\begin{eqnarray}
v_{n+1}&=&uv_{n}+vu_{n}+vE^{-1}(u_{n})+v\,(E^{-1}(u)-u)\,(1-E)^{-1}{v_{n}
\over v},\\
 u_{n+1}&=& E(v_{n})+u u_{n}+v(1-E)^{-1}\, {v_{n} \over
v}-E(v)(1-E)^{-1}E{v_{n} \over v}.
\end{eqnarray}

\noindent From this recursion relation  the recursion operator of
the hierarchy follows

\vspace{0.3cm}

\noindent {\bf Example 7}.\, {\bf Four-field system on time
scale.} We give two examples which are
studied in \cite{blaz1} for the case ${\mathbb T}={\mathbb Z}.$\\
{\bf 1}.\, Let $k=0$ and $\alpha=-2$ and
\begin{equation}
L={\mathcal E}^2+w\,{\mathcal E}+v+u\,{\mathcal
E}^{-1}+p\,{\mathcal E}^{-2}.
\end{equation}
Then we get the four-filed equations

\begin{eqnarray}
\ell=1~~~{dp \over dt_{1}}&=&vp-pE^{-2}(v), \\
{du \over dt_{1}}&=&vu+wE(p)-pE^{-2}(w)-uE^{-1}(v), \\
{dv \over dt_{1}}&=& w E(u)+E^2(p)-uE^{-1}(w)-p,\\
{dw \over dt_{1}}&=&E^{2}(u)-u.
\end{eqnarray}
{\bf 2}.\, Let $k=1$ and $\alpha=-2$ and
\begin{equation}
L=\bar{q}\,{\mathcal E}^2+\bar{w}\,{\mathcal
E}+\bar{v}+\bar{u}\,{\mathcal E}^{-1}+{\mathcal E}^{-2}.
\end{equation}
Then we get another four-filed equations
\begin{eqnarray}
\ell=1~~~{d \bar{u} \over dt_{1}}&=& \bar{w}-E^{-2}(\bar{w}),\\
{d \bar{v} \over dt_{1}}&=& \bar{w}E(\bar{u})+\bar{q}-E^{-2}(\bar{q})-
\bar{u}E^{-1}(\bar{w}),\\
{d \bar{w} \over dt_{1}}&=& \bar{w}E(\bar{v})+\bar{q}E^2(\bar{u})-
\bar{u}E^{-1}(\bar{q})-\bar{v}\bar{w},\\
{d \bar{q} \over dt_{1}}&=&\bar{q}E^{2}(\bar{v})-\bar{v}\bar{q}.
\end{eqnarray}

\vspace{0.3cm}

So far we considered  the hierarchies coming from  Proposition 19
with integer powers of the  Lax operators. Now we consider the
rational powers of the Lax operator.

\vspace{0.3cm}

\noindent {\bf Proposition 22}.\, {\it Let

\begin{equation}
L=w\,{\mathcal E}^{N}+u_{N-1}\, {\mathcal E}^{N-1}+ \cdots +u_{0},
\end{equation}
where $w(x)$ is a function of $x$ which is not a dynamical
variable $dw/dt=0$, $u_{i}~, i=0,1, \cdots, N-1$ are functions of
$t$ and $x \in {\mathbb T}$. Then
\begin{equation}
{dL \over dt_{n}}=[(L^{n/N})_{ \ge 0}, L],~~~n=1,2, \cdots
\label{l17}
\end{equation}
produces hierarchies of integrable systems. Here $n$ is a positive
integer not divisible by $N$. Furthermore the function $u_{0}$ is
also not dynamical, i.e., $u_{0}=u_{0}(x)$, not depending on $t$.}

\vspace{0.3cm}

\noindent {\bf Corollary 23 }. {\it When $N=2$ and $w={1 \over
\mu}E({1 \over \mu})$ then the $\Delta$-KdV Lax operator
(\ref{lax55}) reduces to the above form with

\begin{eqnarray}
u_{0}&=&-{v \over \mu}+{1 \over \mu^2}+u,\\
u_{1}&=&-{1 \over \mu}[E({1 \over \mu})+{1 \over \mu}]+{v
\over\mu}.
\end{eqnarray}
Hence in part (2) of Example 5  we have a single equation with
$v=-\mu u_{0}+{1 \over \mu}+\mu u$.}

\vspace{0.3cm}

 \noindent
In the following example we  study the $N=2$ case  in more detail.

\vspace{0.3cm}

\noindent
 {\bf Example 8}.{\bf KdV on discrete time scales}. Let

\begin{equation}
L=wE(w)\,{\mathcal E}^{2}+u\, {\mathcal E}+v.
\end{equation}
Then

\begin{equation}
L^{1/2}=w\,{\mathcal E}+\alpha_{0}+\alpha_{1}\, {\mathcal
E}^{-1}+\alpha_{2} \, {\mathcal E}^{-2}+ \cdots
\end{equation}
where first three $\alpha_{i}$ are given as
\begin{eqnarray}
w\,(E(\alpha_{0})+\alpha_{0})&=&u, \\
w\,E(\alpha_{1})+E^{-1}(w)\,\alpha_{1}&=&v-(\alpha_{0})^2, \\
w\,E(\alpha_{2})+E^{-2}(w)\, \alpha_{2}&=&-{\alpha_{1}\,E^{-1}(u)
\over E^{-1}(w)}.
\end{eqnarray}
Then we calculate $L^{3/2}$ by
\begin{equation}
L^{3/2}=wE(w)E^2(w)\,{\mathcal E}^3+p_{2}\, {\mathcal E}^2+p_{1}\,
{\mathcal E}+p_{0}+ \mbox{negative powers of}~ {\mathcal E}
\end{equation}
where
\begin{eqnarray}
p_{2}&=&E(w)[w\,E^2(\alpha_{0})+u],\\
p_{1}&=&w E(w)\,E^2(\alpha_{1})+u E(\alpha_{0})+wv, \\
p_{0}&=&wE(w)E^2(\alpha_{2})+u E(\alpha_{1})+v\alpha_{0},\\
&=&wE^{-1}(w)[E^{-1}+E(w)E]^{-1}\,(E(\alpha_{1}){u \over w})+v
(1+E)^{-1}{u \over w}.
\end{eqnarray}
Then (\ref{l17}) with $N=2$ produces a hierarchy of evolution
equations. It turns out that $v$ becomes a constant in the whole
hierarchy. We give the first two members of the hierarchy (for
$n=1$ and $n=3$):

\begin{eqnarray}
u_{t_{1}}&=&u(1-E)(1+E)^{-1}\,u ,\\
u_{t_{3}}&=&u\,(1-E)\,p_{0},
\end{eqnarray}
where $p_{0}$ is given above. The next members of the hierarchy
can be found by taking $n=5$ in (\ref{l17}) or by applying the
recursion operator ${\mathcal R}$ to $u_{t_{3}}$. \noindent For
${\mathbb T}={\mathbb K}_{q}$ and  $w=1$ the above hierarchy and
its Hamilton formulation were given by Frenkel \cite{fren}. The
recursion operator of this hierarchy with $w=1$ can be found by
using (\ref{rec}) with $R_{n}=\alpha_{n} {\mathcal E}+\beta_{n}$.
We find that

\begin{eqnarray}
(E^2-1)\, \alpha_{n}&=&E^2 (u_{n}),\\
(E^2-1)\, \beta_{n}&=&uE(u_{n})+E(u)\, \alpha_{n}-u\,
E(\alpha_{n})
\end{eqnarray}
and the equation which determines the recursion operator is
\begin{equation}
u_{n+1}=v u_{n}-u (E-1)\, \beta_{n},~~~n=0,1,2, \cdots .
\end{equation}
We find that
\begin{equation}
{\mathcal R}=v-u(E+1)^{-1}\,[-u+E(u)E](E^2-1)^{-1}\,E.
\end{equation}

\vspace{0.3cm}

 When the Lax operator is of degree one  and has  an infinite
power series in  operator ${\mathcal E}^{-1}$ the corresponding
system is called the KP hierarchy.

\vspace{0.3cm}

\noindent {\bf Proposition 24}.\, {\it Let
\begin{equation}
L={\mathcal E}+u_{0}+u_{1}\, {\mathcal E}^{-1}+u_{2}\, {\mathcal
E}^{-2}+\cdots .
\end{equation}
Then
\begin{equation}
{dL \over dt_{n}}=[(L^{n})_{ \ge 0}, L], ~~~ n=1,2,\cdots ,
\end{equation}
produces the following hierarchy

\begin{eqnarray}
n=1~~~{du_{0} \over dt_{1}}&=&(E-1)u_{1}, \\
{du_{1} \over dt_{1}}&=&(E-1)u_{2}+u_{1}[u_{0}-E^{-1}(u_{0})],\\
&&............................. \nonumber\\
{du_{k} \over
dt_{1}}&=&(E-1)\,u_{k+1}+u_{k}\,[u_{0}-E^{-k}(u_{0})],~~~k=0,1,
\cdots .
\end{eqnarray}

\begin{eqnarray}
n=2~~~{du_{0} \over dt_{2}}&=&(E^2-1)u_{2}-u_{1}E^{-1}(E+1)\,u_{0} \nonumber\\
&&+E(u_{1})(E(u_{0})+u_{0}), \\
{du_{1} \over
dt_{2}}&=&(E^2-1)\,u_{3}+\alpha_{1}E(u_{2})-u_{2}E^{-2}(\alpha_{1})
\nonumber \\
&&+\alpha_{0}u_{1}-u_{2} E^{-1}(\alpha_{0}),\\
 &&............................. ,\nonumber\\
{du_{k} \over
dt_{2}}&=&(E^2-1)u_{k+2}+\alpha_{1}E(u_{k+1})-u_{k+1}E^{-k-1}(\alpha_{1})
\nonumber\\
&& +\alpha_{0}u_{k}-u_{k}E^{-k}(\alpha_{0}),  ~~~ k=0,1, \cdots,
\end{eqnarray}}
where $\alpha_{0}=(E+1)u_{1}+(u_{0})^2$ and
$\alpha_{1}=(E+1)u_{0}$.
 The case  ${\mathbb T}={\mathbb Z}$ of this hierarchy is
discussed in \cite{blaz1} (see also the references therein) and
the case ${\mathbb T}={\mathbb K}_{q}$  is discussed in
\cite{fren} and  \cite{thre}.

\section*{6. Trace Functional and Conservation Laws}

Let ${\mathbb T}$ be a regular time scale and $\Lambda$ be the
algebra of pseudo delta differential operators. Any operator $F
\in \Lambda$ of order $k$ has the form

\begin{equation}\label{exp1}
F=a_{k}\, \delta^{k}+a_{k-1}\, \delta^{k-1}+ \cdots + a_{1}\,
\delta+a_{0}+a_{-1}\, \delta^{-1}+a_{-2}\, \delta^{-2}+\cdots
\end{equation}
where $a_{\ell}$'s are $\Delta$-smooth functions of $x \in
{\mathbb T}$ (they are also functions of $t \in {\mathbb R}$). The
coefficients $a_{0}$ and $a_{-1}$ we call respectively the {\it
free term} (zero order term) and the {\it  residue } of $F$
associated with its "$\delta$-expansion" (\ref{exp1}) and write

\begin{equation}\label{not1}
\mbox{Free}_{\delta}\, F=a_{0}(x) ~~~ \mbox{and} ~~~
\mbox{Res}_{\delta}\, F=a_{-1}(x).
\end{equation}
In case of regular-discrete time scales ${\mathbb T}$ we have
\begin{equation}\label{delta1}
\delta={1 \over \mu}\, ({\mathcal E}-I)={1 \over \mu}\, {\mathcal
E}-{1 \over \mu}
\end{equation}
and therefore the same operator $F$ can be expanded in series with
respect to the powers of ${\mathcal E}$ of the form

\begin{equation}\label{exp2}
F=b_{k}\, {\mathcal E}^{k}+b_{k-1}\, {\mathcal
E}^{k-1}+\cdots+b_{1}\, {\mathcal E}+b_{0}+b_{-1}\, {\mathcal
E}^{-1}+b_{-2}\, {\mathcal E}^{-2}+ \cdots
\end{equation}
We write
\begin{equation}\label{not2}
\mbox{Free}_{\mathcal E}\, F=b_{0}(x) ~~~ \mbox{and} ~~~
\mbox{Res}_{\mathcal E}\, F=b_{-1}(x).
\end{equation}
Substituting (\ref{delta1}) and

\begin{equation}
\delta^{-1}=({\mathcal E}-I)^{-1}\, \mu=({\mathcal
E}^{-1}+{\mathcal E}^{-2}+\cdots)\, \mu=E^{-1}(\mu)\, {\mathcal
E}^{-1}+E^{-2}(\mu)\, {\mathcal E}^{-2}+ \cdots ,
\end{equation}
into (\ref{exp1}) and taking into account that

\begin{equation}
E^{-1}(\mu)=\mu(\rho(x))=\sigma(\rho(x))-\rho(x)=x-\rho(x)=\nu(x),
\end{equation}
we find that

\begin{equation}\label{res1}
\mbox{Res}_{\mathcal E}\, F=\nu\, \mbox{Res}_{\delta}\, F.
\end{equation}

\vspace{0.3cm}

\noindent {\bf Definition 25.}\, {\it The trace of an operator $F
\in \Lambda $ is defined by

\begin{equation}
\mbox{Tr}(F)=\int_{\mathbb T}\, \mbox{Res}_{\delta}\, \{F(I+\mu
\delta)^{-1} \}\, \nabla x,
\end{equation}
where the nabla integral is defined according to Section 2.}

\vspace{0.3cm}

\noindent {\bf Proposition 26.}\, {\it Let $F$ be given as in
(\ref{exp1}). In case of regular-discrete time scales we have

\begin{equation}
\mbox{Res}_{\delta} \{F(I+\mu \delta)^{-1} \}={ 1 \over \nu(x)}\,
\mbox{Free}_{\mathcal E}\,F
\end{equation}
for $x \in {\mathbb T}\setminus \{x_{*},x^{*} \}$, where
$x_{*}=\min {\mathbb T}$ and $x^{*}=\max {\mathbb T}$. Therefore
in this case

\begin{equation}
\mbox{Tr}(F)=\int_{\mathbb T}\, (\mbox{Free}_{\mathcal E}F){\nabla
x \over \nu(x)}=\sum_{x \in {\mathbb T}}\, b_{0}(x).
\end{equation}}

\vspace{0.3cm}

\noindent {\bf Proof}. Since $I+\mu \delta={\mathcal E}$ we have,
by using (\ref{res1}) and (\ref{exp2}),

\begin{eqnarray}
\nu\, \mbox{Res}_{\delta} \{F(I+\mu \delta )^{-1}
\}=\mbox{Res}_{\mathcal E} \{ F(I+\mu \delta)^{-1}
\}=\mbox{Res}_{\mathcal E}\, (F
{\mathcal E}^{-1}) \nonumber \\
=\mbox{Res}_{\mathcal E}\,(b_{k} {\mathcal E}^{k-1}+ \cdots
+b_{1}+b_{0}\, {\mathcal E}^{-1}+ \cdots
)=b_{0}=\mbox{Free}_{\mathcal E}(F).
\end{eqnarray}

\vspace{0.3cm}

\noindent {\bf Proposition 27}.\, {\it For all $F_{1}, F_{2} \in
\Lambda $

\begin{equation}\label{tr1}
\mbox{Tr}([F_{1},F_{2}])=\mbox{Tr}(F_{1} F_{2}-F_{2} F_{1})=0,
\end{equation}
in other words the pairing $(F_{1},F_{2})=Tr(F_{1} F_{2})$ is
symmetric.}

\vspace{0.3cm}

\noindent We prove (\ref{tr1}) only for particular cases of time
scales ${\mathbb T}$. They indicate a way to the proof in the
general case of regular time scales.

\noindent {\bf (i)}. \, If ${\mathbb T}={\mathbb R}$, then
$\delta=\partial={d \over dx} \cdot$ and $\mu(x)=0$,

\begin{equation}
F=a_{k} \partial^{k}+\cdots +a_{1} \partial+a_{0}+a_{-1}
\partial^{-1}+ \cdots
\end{equation}
and
\begin{equation}
\mbox{Tr}(F)=\int_{\mathbb R}\, \mbox{Res}(F) dx=\int_{\mathbb
R}\, a_{-1}(x) dx .
\end{equation}
It is well known that (for example, see \cite{blaz1}) for such
functional $Tr(F)$ the statement (\ref{tr1}) holds.

\vspace{0.3cm}

\noindent {\bf (ii)}.\,Let ${\mathbb T}$ be a regular-discrete
time scale. Then by Proposition 26 we have
\begin{equation}\label{tr2}
\mbox{Tr}([F_{1},F_{2}])=\int_{\mathbb T}\mbox{Res}_{\delta}
\{[F_{1},F_{2}] (I+\mu \delta)^{-1} \} \nabla x=\int_{\mathbb T}\,
(\mbox{Free}_{\mathcal E} [F_{1},F_{2}]) {\nabla x \over
\nu(x)}=0.
\end{equation}
It is enough to check (\ref{tr2}) for monomials $F_{1}=A {\mathcal
E}^{k}$ and $F_{2}=B{\mathcal E}^{\ell}$. By the use of the
property (\ref{epsil}) of ${\mathcal E}$ we have
\begin{equation}
F_{1}F_{2}=A(E^{k}\,B)\, {\mathcal E}^{k+\ell} ~~~
\mbox{and}~~~F_{2}F_{1}=B(E^{\ell}\,A).
\end{equation}
Therefore $Free_{\mathcal E}[F_{1},F_{2}]$ is either zero or

\begin{eqnarray}
&&\mbox{Free}_{\mathcal E}[F_{1},F_{2}]=A(E^{k} B)-B
(E^{-k}\,A)=(E^k-I)(E^{-k}A)B \nonumber\\
&&=(I-E^{-1})\,(E^{k}+E^{k-1}+\cdots+E+I)(E^{-k}A)B=\nu(x)[\Phi(A,B)]^{\nabla},
\nonumber
\end{eqnarray}
where $\Phi(A,B)=(E^{k}+E^{k-1}+\dots+E+I)(E^{-k}A)B$. Hence
\begin{equation}
\int_{\mathbb T}\, \mbox{Free}_{\mathcal E} [F_{1},F_{2}] \,
{\nabla x \over \nu(x)}=\int_{\mathbb T}[\Phi(A,B)]^{\nabla}
\nabla x=\Phi(A,B)|^{x^{*}}_{x_{*}}=0
\end{equation}
so that (\ref{tr1}) is proved for regular-discrete time scales.

\vspace{0.3cm}

\noindent {\bf (iii)}.\, Let ${\mathbb T}$ be a mixed time scale,
say, of the form ${\mathbb T}=(-\infty, 0) \cup {\mathbb K}_{q}$,
where $(-\infty,0)$ denotes the real line interval. Then for any
$F_{1},F_{2} \in \Lambda$ we have, taking into account Proposition
26, that
\begin{eqnarray}
\mbox{Tr}([F_{1},F_{2}])=\int_{\mathbb T}\, \mbox{Res}_{\delta}\,
\{[F_{1},F_{2}](I+\mu \delta)^{-1} \} \nabla x \nonumber \\
=\int_{-\infty}^{0}\,(\mbox{Res}_{\partial}([F_{1},F_{2}])
dx+\int_{{\mathbb K}_{q}}\, (\mbox{Free}_{\mathcal
E}\,[F_{1},F_{2}]) {\nabla x \over \nu(x)}.
\end{eqnarray}
Take for instance $F_{1}=A \delta$ and $F_{2}=B \delta^{-1}$. Then

\begin{eqnarray}
&&\mbox{Res}_{\partial}\,[F_{1},F_{2}]=\mbox{Res}_{\partial}\,[A
\partial, B
{\partial}^{-1}]=AB^{\prime}+A^{\prime}B=(AB)^{\prime}, \nonumber\\
&&\mbox{Free}_{\mathcal E} [F_{1},F_{2}]=\mbox{Free}_{\mathcal
E}\,[{A \over \mu}\,({\mathcal E}-I),B({\mathcal E}-I)^{-1}\, \mu]
\nonumber\\
 &&=AE(B)-BE^{-1}(A)
\end{eqnarray}
Therefore

\begin{eqnarray}
\int_{-\infty}^{0}\,(\mbox{Res}_{\partial} [F_{1},F_{2}])
dx=\int_{-\infty}^{0}\,(AB)^{\prime} dx=A(0^{-})B(0^{-}), \\
\int_{{\mathbb K}_{q}}\,\mbox{Free}_{\mathcal E}
[F_{1},F_{2}]\,{\nabla x \over \nu(x)}=\int_{{\mathbb
K}_{q}}[AE(B)-BE^{-1}(A)]\,{\nabla x \over \nu(x)} \nonumber\\
=\sum_{x \in {\mathbb
K}_{q}}\,[A(x)B(qx)-A(q^{-1}x)B(x)]=-A(0^{+})B(0^{+}).
\end{eqnarray}
Hence
\begin{equation}
\mbox{Tr}([F_{1},F_{2}])=A(0^{-})B(0^{-})-A(0^{+})B(0^{+})=0,
\end{equation}
where $A$ and $B$ are $\Delta$-smooth functions on ${\mathbb T}$
and  hence they are continuous  at $x=0$.

\vspace{0.3cm}

\noindent {\bf Proposition 28}\,{\it Equation (\ref{lax5}) implies
that
\begin{equation}
{d  \over dt_{n}}L^{k}=[A_{n},L^{k}], ~~~~ A_{n}=(L^{n \over N})_{
\ge 0},
\end{equation}
for all $k={\ell \over N}$, where $\ell$ is any positive integer.}

\vspace{0.3cm}

\noindent Propositions 27 and 28 imply the next Proposition

\vspace{0.3cm}

\noindent {\bf Proposition 29}.\,{\it  For all $\ell=0,1,\cdots $
the functionals

\begin{equation}\label{con1}
H_{\ell}=\mbox{Tr} (L^{\ell \over N}),
\end{equation}
are common constants of motion for the hierarchy (\ref{lax5}) and
(\ref{lax19}).}

\vspace{0.3cm} \noindent
 Note that in proof of Proposition 29 it is, in particular, used
 the fact that the flows (vector fields) defined by the different
 members of the hierarchy all commute with each other (see
 \cite{diki}, \cite{will}).
\section*{8. Conclusion}

We have developed the Gel'fand-Dikii approach to time scales. So
far the integrable systems were studied on ${\mathbb R}, {\mathbb
Z}$ or on ${\mathbb K}_{q}$. Here we gave a unified and extended
approach. In particular cases when ${\mathbb T}={\mathbb R},
{\mathbb Z}, {\mathbb K}_{q}$ we find  several examples of the
integrable systems. We developed the algebra of $\Delta$-pseudo
differential and ${\mathcal E}$-shift operators. We established
the GD formalism on these algebras and introduced several Lax
representations on these algebras. All these Lax representations
are straightforward generalizations of the Lax representations on
pseudo differential algebras of integrable systems on ${\mathbb
R}$ and the Lax representations of the algebra of shift operators
on ${\mathbb Z}$. The Burgers and KdV hierarchies on time scales
that we found are the special cases of these Lax representations.
We also generalized the Frenkel KdV hierarchy introduced on
${\mathbb K}_{q}$ to arbitrary discrete time scales. We
constructed the recursion operators of each example considered in
this paper and gave a way of constructing the constants of motions
by introducing an appropriate trace form on time scales.

In this work we did not consider the r-Matrix construction and the
Hamiltonian formulation of integrable systems on time scales. The
trace form on a general time scale needs a little care. Such a
work is in progress and will be communicated in a separate paper.

\section*{Acknowledgments}

This work is partially  supported by the Turkish Academy of
Sciences and by the Scientific and Technical Research Council of
Turkey.

\vspace{1cm}
\newpage
\section*{APPENDIX: Recursion operators of four field systems}
 We give the recursion operator of the Four-field systems on time
scale which are studied in example 7.

\vspace{0.3cm}

\noindent\textbf{1.} For the case $k=0$, $\alpha=-2$, we obtain
the recursion relation between the $n+1$-th and $n$-th elements of
the hierarchy as follows:
\begin{eqnarray}
w_{n+1}=&w(E+1)^{-1}E(v_{n})+E(v)(E^{2}-1)^{-1}E^{2}(w_{n})
-v(E^{2}-1)^{-1}w_{n}+\nonumber\\&w(E+1)^{-1}(1-E)w(E^{2}-1)^{-1}E(w_{n})+E^{2}(u_{n})+(1-E^{2})\eta_{n}
\label{a1},\\
v_{n+1}=&w
E(u_{n})+vv_{n}-u(E^{2}-1)^{-1}E^{-1}(w_{n})+E(u)(E^{2}-1)^{-1}E^{2}(w_{n})+\nonumber \\
&(1-E^{2})p(1-E^{2})^{-1}
\frac{p_{n}}{p}+E^{2}(p_{n})+(E^{-1}(w)-wE) \eta_{n}\label{a2},\\
u_{n+1}=&wE(p_{n})+u(E+1)^{-1}(E+1+E^{-1})v_{n}-p(E^{2}-1)^{-1}E^{-2}(w_{n})+\nonumber\\
&u(E+1)^{-1}(E-1)E^{-1}(w)(E^{2}-1)^{-1}w_{n}+E(p)(E^{2}-1)^{-1}E^{2}(w_{n})+\nonumber\\
&(E^{-2}(w)-w
E)p(1-E^{2})^{-1}\frac{p_{n}}{p}+vu_{n}+(E^{-1}(v)-v)\eta_{n}\label{a3},\\
p_{n+1}=&uE^{-1}(u_{n})+p(1+E^{-2})v_{n}+p(E^{-1}-E^{-2})w(E^{2}-1)^{-1}E(w_{n})
+\nonumber \\
&(E^{-2}(v)-v)p(1-E^{2})^{-1}\frac{p_{n}}{p}+vp_{n}+(E^{-1}(u)-uE^{-1})\eta_{n},
\label{a4}
\end{eqnarray}

\vspace{0.3cm}

\noindent where
\begin{eqnarray}
\eta_{n}=&(E^{2}(p)-E(p)E^{2})^{-1}[E^{2}(u)E(p_{n})+E^{2}(p)u_{n}+
(uE^{2}(p)-E^{2}(u)E(p))\nonumber\\
&(1-E^{2})^{-1}E^{2}(\frac{p_{n}}{p})].\nonumber
\end{eqnarray}

\vspace{0.3cm}

\noindent\textbf{2.} For the case $k=1$, $\alpha=-2$, the
recursion relation between the $n+1$-th and $n$-th elements of the
hierarchy is given by

\vspace{0.3cm}

\begin{eqnarray}
\bar{u}_{n+1}=&E^{-2}(\bar{w}_{n})+\bar{u}(1+E)^{-1}(E-1)\bar{u}(1-E^{2})^{-1}E(\bar{u}_{n})
+\bar{v}\bar{u}_{n}+\bar{u}(1+E)^{-1}\bar{v}_{n}+\nonumber\\&(E^{-1}(\bar{v})-\bar{v})
(1-E^{2})^{-1}\bar{u}_{n}+(1-E^{-2})\bar{\zeta}_{n}\label{b1},\\
\bar{w}_{n+1}=&\bar{u}E^{-1}(\bar{q}_{n})+\bar{w}(1+E)^{-1}(E^{2}+E+1)\bar{v}_{n}+
(E^{-1}(\bar{q})-\bar{q}E^{4})(1-E^{2})^{-1}\bar{u}_{n}+
\nonumber\\&\bar{w}(1+E)^{-1}(1-E)E(\bar{u})(1-E^{2})^{-1}E^{2}(\bar{u}_{n})+\bar{v}\bar{w}_{n}
+(E(\bar{v})-\bar{v})\bar{\zeta}_{n}+\nonumber\\&(E^{2}(\bar{u})-\bar{u}E^{-1})
\bar{q}(E^{2}-1)^{-1}E^{2}(\frac{\bar{q}_{n}}{\bar{q}})\label{b2},\\
\bar{q}_{n+1}=&\bar{v}\bar{q}_{n}+\bar{w}E(\bar{w}_{n})+\bar{q}(1+E^{2})\bar{v}_{n}
+(E^{2}(\bar{v})-\bar{v})\bar{q}(E^{2}-1)^{-1}E^{2}(\frac{\bar{q}_{n}}{\bar{q}})+\nonumber\\&\bar{q}(1+E)^{-1}(1-E^{2})E(\bar{u})
(1-E^{2})^{-1}E^{2}(\bar{u}_{n})+
(E(\bar{w})-\bar{w} E)\bar{\zeta}_{n}\label{b3},\\
\bar{v}_{n+1}=&E^{-2}(\bar{q}_{n})+\bar{u}E^{-1}(\bar{w}_{n})+
(E^{-1}(\bar{w})-\bar{w}E^{3})(1-E^{2})^{-1}\bar{u}_{n}+\bar{v}\bar{v}_{n}+\nonumber\\&(E(\bar{u})-\bar{u}E^{-1})
\bar{\zeta}_{n}+
(1-E^{2})\bar{q}(E^{2}-1)^{-1}E^{2}(\frac{\bar{q}_{n}}{\bar{q}}).\label{b4},
\end{eqnarray}

\noindent where

\begin{equation}
\bar{\zeta}_{n}=(\bar{q}E^{2}-E(\bar{q}))^{-1}[\bar{q}E^{2}(\bar{w}_{n})
+(E^{2}(\bar{w})\bar{q}E-
\bar{w}E(\bar{q}))(E^{2}-1)^{-1}E(\frac{\bar{q}_{n}}{\bar{q}})].\nonumber
\end{equation}

\noindent From the recursion relations obtained in both cases, we
can construct the recursion operators of the hierarchies.

\end{document}